\documentclass[12pt]{article}
\usepackage{amsmath,amssymb}

\newcommand{\ncm}{\newcommand}
 \ncm{\R}{\mathbb{R}}
 \ncm{\C}{\mathbb{C}}
 \ncm{\Q}{\mathbb{Q}}
 \ncm{\N}{\mathbb{N}}
 \ncm{\Ad}{\mbox{\rm Ad}}
 \ncm{\ad}{{\rm ad}}
 \ncm{\Tr}{{\rm Tr}}

 \ncm{\Aut}{\mbox{\rm Aut}}
 \ncm{\Sp}{{\rm Sp}}
 \ncm{\supp}{{\rm supp}}
 \ncm{\Ker}{{\rm Ker}}
 \ncm{\id}{{\rm id}}
 \ncm{\ra}{\rightarrow}
 \ncm{\cstar}{C$^*$-algebra}
 \ncm{\M}{{\mathcal{M}}}
 \ncm{\Ex}{{\mathcal Ex}}
 \ncm{\E}{{\mathcal E}}
 \ncm{\G}{{\mathcal G}}
 \ncm{\F}{{\mathcal F}}
 \ncm{\B}{{\mathcal B}}
 \ncm{\T}{\mathbb{T}}
 \ncm{\V}{{\mathcal V}}
 \ncm{\cL}{{\mathcal L}}
 \ncm{\cP}{{\mathcal P}}
 \ncm{\cS}{{\mathcal S}}
 \ncm{\I}{{\mathcal I}}
 \ncm{\K}{{\mathcal K}}
 \ncm{\U}{{\mathcal U}}
 \ncm{\Hil}{{\mathcal H}}

 \ncm{\Z}{{\mathbb{Z}}}
 \ncm{\eps}{\epsilon}
 \ncm{\ran}{{\rangle}}
 \ncm{\lan}{{\langle}}
 \ncm{\addednote}{\footnote}

 \newtheorem{theo}{Theorem}[section]

\newtheorem{prop}[theo]{Proposition}
\newtheorem{remark}[theo]{Remark}
\newtheorem{definition}[theo]{Definition}
\newtheorem{example}[theo]{Example}

\title{A mathematical model for measurement}
\author{Aki Kishimoto\footnote{ E-mail: akiksmt@r3.ucom.ne.jp; Retired from Hokkaido University}}
\date{}
\begin{document}

\maketitle

\begin{abstract}
We will give a new model for measurement of a quantum system on a Hilbert space such that the measuring apparatus is described by a unital separable non-type I nuclear simple C$^*$-algebra equipped with a certain unital endomorphism and a pure state. An interaction between the quantum system and the apparatus is specified by a unitary associated with the combined system as before. Magnifying to the classical level some aspects of the quantum system so captured in the apparatus is explicitly done by applying the endomorphism; then the resulting state is the superposition of {\em phases} with weights. Nature will then choose each phase according to the probability prescribed by the weights just as does one when multiple phases appear as in phase transition. Thus in our model state-reduction (or collapse of the wave function) is a primary event; whether this corresponds to the measurement of an observable or which one if it does is another matter.
\end{abstract}

\section{Introduction}
A C$^*$-algebraic approach was proposed for understanding certain quantum phenomena more than half a century ago. As notable examples we point out a full-blown theory for equilibrium states in quantum statistical mechanics (cf. \cite{BR}) and an axiomatic theory of quantum field theory (cf. \cite{H}). What we realize from this perspective is that believing in the {\em universal wave function} which is purported to describe the universe may not be untenable; or when we deal with a quantum system which is so enormous to allow macroscopically distinguishable states the right framework may be a C$^*$-algebra formalism; at least a first approximation to the description of such a system is not a quantum mechanics on a Hilbert space, but rather a dynamical system based on a certain C$^*$-algebra of obervables. This is an insight which both derived and was derived by the introduction of a concept of C$^*$-algebra to mathematical formulation of quantum mechanics.

The time development of quantum mechanics on a Hilbert space is described by unitary operators and hence is reversible. Namely in the Heisenberg picture an observable (or a self-adjoint operator on the Hilbert space) $Q$ is mapped into $U^*QU$ under a passage of time where $U$ is an appropriate unitary operator. Departing from quantum mechanics as an operator theory on Hilbert spaces as described in \cite{vN} we realize the time development on C$^*$-algebras (instead of the algebra of all bounded operators) need not be reversible (though preserving the algebraic relations) even if we assume that the time development can be approximated by automorphisms  induced by unitaries as should be. (See \cite{K03} for an abundance of examples of such endomorphisms.) Thus some irreversibility is intrinsic in the C$^*$-algebraic formalism of quantum mechanics without invoking an argument of {\em disturbances} from the ambient system for example. (But note that in the aforementioned theories only automorphisms are considered as time developments.) We shall apply this approach to the measuring process, which is an irreversible process of the combined system of a quantum system based on a Hilbert space and a measuring apparatus which must be huge enough to allow macroscopically different states.

Before going into details let us explain our point of view in more general terms.

A state $\omega$ on a C$^*$-algebra $A$ is a continuous linear functional on $A$ such that $\omega(x^*x)\geq0,\ x\in A$ and $\|\omega\|=1$. In general $\omega$ is supposed to represent a state of the system represented by $A$, hence it bears the name of state. A (non-degenerate) representation $\pi$ of $A$ is a linear map of $A$ into the bounded operators $\B(\Hil_\pi)$ on a Hilbert space $\Hil_\pi$ such that $\pi(xy)=\pi(x)\pi(y)$ and $\pi(x)^*=\pi(x^*)$ for all $x,y\in A$ and the linear span of $\pi(A)\Hil_\pi$ is dense in $\Hil_\pi$. Two representations $\pi_1$ and $\pi_2$ of $A$ are called  quasi-equivalent if the kernels of $\pi_1$ and $\pi_2$ are equal and the map $\pi_1(x)\mapsto \pi_2(x)$ of $\pi_1(A)$ onto $\pi_2(A)$ is continuous in the weak operator topology. Given a state $\omega$ there is a  representation $\pi_\omega$ of $A$, called the GNS representation; we recover $\omega$ by $\omega(x)=\lan \pi_\omega(x)\Omega_\omega,\Omega_\omega\ran,\ x\in A$ for some unit vector $\Omega_\omega$ in the representation space $\Hil_\omega$. By imposing the condition that $\pi_\omega(A)\Omega$ is dense in $\Hil_\omega$ the triple $(\pi_\omega,\Hil_\omega,\Omega)$ is essentially uniquely determined by $\omega$. Let $\M$ denote the weak closure of $\pi_\omega(A)$ in $\B(\Hil_\omega)$ and let $\M'=\{ Q\in \B(\Hil_\omega)\ |\ \forall T\in\M\ QT=TQ \}$, which is called the commutant of $\M$. We note that $\M$, $\M'$, and $\B(\Hil_\omega)$ are all von Neumann algebras (in the sense that they are weakly closed $*$-closed algebras of bounded operators on a Hilbert space) and that $\M=(\pi_\omega(A)')'\equiv \pi_\omega(A)''$.  A normal state $\phi$ on $\M$ is a state on $\M$ obtained as $\phi(Q)=\sum_i \lan Q\xi_i,\xi_i\ran, \ Q\in\M$ where $\xi_i\in \Hil_\pi$ for $i=1,2,\ldots$ and $\sum_i\|\xi_i\|^2=1$. We call the state $\phi\pi_\omega$ of $A$ a state {\em affiliated with} $\pi_\omega$; in particular $\omega$ is affiliated with $\pi_\omega$. We call $\M$ a factor and $\omega$ a factorial state if $\M\cap \M'=\C1$.

Let $\omega$ be a factorial state of $A$; then any two states affiliated with $\pi_ \omega$ are not considered as being macroscopically distinguished. One reasoning for this is as follows: For a bounded sequence $(x_n)$ in $A$ with $\|x_ny-yx_n\|\ra0,\ y\in A$ we choose a subset $(x_\iota)$ of $(x_n)$ such that $\pi_\omega(x_\iota)$ weakly converges on $\Hil_\omega$ where the limit must be a constant multiple of the identity. Then it follows that $\lim_\iota\omega(x_\iota)=\lim_\iota\phi(x_\iota)$ for any state $\phi$ affiliated with $\pi_\omega$; thus it follows that $\lim_n(\omega(x_n)-\phi(x_n))=0$. This fact may be interpreted as showing that $\omega$ and $\phi$ are not distinguishable by observables obtained by {\em averaging process}; thus they are not macroscopically distinguishable.

Our view on a C$^*$-algebraic approach to physical systems are as follows.\footnote{Which I believe was generally held when the C$^*$-algebra theory was first introduced into mathematical physics but seems to have been forgotten since.} Let $A$ be a C$^*$-algebra describing a physical system and let $FS(A)$ denote the set of factorial states of $A$ and $FR(A)$ be the set of quasi-equivalent classes of factorial representations of $A$. There is a map of $FS(A)$ onto $FR(A)$ by mapping $\omega$ to the quasi-equivalent class $[\pi_\omega]$  of $\pi_\omega$. Note that all states affiliated with $\pi_\omega$ map to $[\pi_\omega]$. An element of $FR(A)$ is regarded as a phase (or pure phase) macroscopically different (or disjoint in mathematical jargon) while an element of $FS(A)$ is regarded as a {\em real} state of the system which may be deciphered by going through measurement process. If the time development is specified by an automorphism $\alpha$ of $A$ this induces a closed dynamics on $FR(A)$ by $[\pi]\mapsto [\pi\alpha]$ and we do not have to consult $FS(A)$ for understanding this dynamics. If it is specified by an endomorphism $\alpha$ then $\pi\alpha$ may not be factorial for a factorial representation $\pi$ and $\pi\alpha$ is uniquely decomposed into factorial representations in a sense, say e.g., $\pi\alpha=\pi_1\oplus \pi_2 \oplus\cdots\oplus\pi_n$, where $\pi_i$'s are mutually disjoint factorial representations.  (In general it is given as a direct integral; see \cite{Sak}.) Thus $[\pi]$ is mapped into $[\pi\alpha]$ which is expected to be realized as one of $[\pi_1], [\pi_2],\ldots,[\pi_n]$ in the real world. A general interpretation of quantum mechanics depicts the probability with which each $[\pi_i]$ is realized; but to compute these probabilities we have to go back to the state in $FS(A)$ behind $[\pi]$. (If $\omega$ is a state of the system which maps into $[\pi]$ and $E_i$ the central projection corresponding to $\pi_i$ then $[\pi_i]$ is realized with the probability $\omega(E_i)$ where $\omega$ is regarded as a state on $\pi(A)''$ and the state hidden behind $[\pi_i]$ is $\omega(E_i\gamma(\,\cdot\,))/\omega(E_i)$ if $\omega(E_i)\not=0$.) Namely the dynamics on $FR(A)$ is not closed by itself in general; the fact that the dynamics on $FR(A)$ is not closed is a window through which we can glimpse a real nature of the state giving rise to an element of $FR(A)$. Measurement is, in our view, to utilize this irreversibility of dynamics which can be naturally realized as endomorphisms of C$^*$-algebras. (If $A$ is a commutative C$^*$-algebra, say, the continuous functions on a compact topological space $X$, then $FS(A)$, $FR(A)$, and $X$ are identified with each other; so no measurement is needed on states. If $A$ is the compact operators on a Hilbert space, then $FR(A)$ is a singleton consisting of the class of the identity representation while $FS(A)$ is all the states of $A$ which are identified with all the positive trace-class operators with trace 1; so in this case we have no information whatever on the state of the system without performing measurement.\footnote{An automorphism which induces the identity map on $FR(A)$ is {\em universally weakly inner}. In particular, if $\alpha$ is inner, i.e., $\alpha=\Ad\,u$ for some unitary or unitary multiplier $u$ of $A$ then $\alpha$ induces the identity map on $FR(A)$. The converse holds if $A$ is a simple separable C$^*$-algebra. See \cite{K81}.})

Measuring a certain quantum system may be imagined by a layman as follows: There is a {\em black box} with a switch and a display, which is fed a quantum system  that is to be measured and is prepared in a specific way. After initializing the black box we feed the quantum system into it, turn on the switch, and wait for a short period of time. Then the display will show an integer which may differ on each trial, i.e., the whole quantum system represented by the black box consisting of the quantum system and the apparatus interacting with it (presumably excluding the classical part leading up to the display, a part which recognizes a macroscopic status of the whole quantum system) starting from the same state will evolve into one of a variety of states which are macroscopically distinguishable, i.e., the initial state of the whole quantum system evolves into a mixed state. If the apparatus is represented by a C$^*$-algebra $A$ and the quantum system by $\K$, the C$^*$-algebra of compact operators on a Hilbert space $\Hil$,\footnote{The algebra $\B(\Hil)$ is recovered as the multiplier algebra of $\K$.} then the whole quantum system is represented by the tensor product $\K\otimes A$. For this kind of evolution to be possible the process occurring in the black box must be represented by a proper endomorphism of $\K\otimes A$, a non-surjective isomorphism of $\K\otimes A$ into $\K\otimes A$; to allow such a non-trivial endomorphism $A$ must not be of type I.\footnote{$A$ is of type I if and only if $[\pi_1]=[\pi_2]$ is equivalent to $\Ker(\pi_1)=\Ker(\pi_2)$ in $FR(A)$ (or in $IR(A)$, the equivalence classes of irreducible representations) for separable C$^*$-algebras. See Theorem 6.8.7 of \cite{Ped}.} Thus if we adopt a C$^*$-algebra framework for describing a measuring process we are necessarily led to incorporating  non-type I C$^*$-algebras.
%And here we adopt a convention that e.g., if a state on $A\otimes\K$ is given by $\omega=\sum_i\lambda_i\omega_i$ with $\lambda_i>0$ where $\omega_i$ are mutually {\em disjoint} (or macroscopically different) pure states and $\sum_i\lambda_i=1$, $\omega$ is a mixed state and represents a mixture of $\omega_i$'s where $\omega_i$ occurs with the probability $\lambda_i$. In otherwise each {\em realization} of $\omega$ in this world is $\omega_i$ with probability $\lambda_i$; a probabilistic nature of quantum mechanics naturally comes in.  Which I believe is accepted in general and certainly more acceptable than {\em the collapse of wave-function} (see \cite{Hep} for the same idea).
The departure from $\K$ (or the quantum theory built on Hilbert spaces) and the usage of endomorphisms do not seem to be explicitly mentioned in the literature, which we will exploit here to some details.\footnote{Some people may say this is the same as adopting {\em decoherence} for explaining the measurement problem; but note that we do not have to introduce the effect of {\em environment} or such for deriving this formalism, which yields more likely (non-endomorphic) completely positive maps instead of endomorphisms. Also note that a semi-flow of endomorphisms on a C$^*$-algebra gives us a new kind of dynamics, a non-deterministic dynamics for each {\em realization}, which is more susceptible of many-world interpretation than a flow of automorphisms and that if the C$^*$-algebra is commutative as in the classical case then we will not get anything new; the semi-flow is just deterministic. Thus one notices that the C$^*$-algebraic formalism of quantum system naturally induces non-deterministic dynamics. }

\small
This endeavor was prompted by Professor M. Ozawa's lectures on his version of Heisenberg's uncertainty principle which nevertheless bewildered me about the idea of measuring process itself, undoubtedly due to my ignorance, at the conference held in February 2013 organized by Professor T. Teruya for operator algebraists. I want to express my thanks to both of them for this opportunity and to Reiko, my wife, who accompanied me for this trip to Kyoto and then an expedition to Furano in March where an inceptive idea to the present note was conceived on a trail, for her unfailing company and patience in listening to my gibberish. I also want to extend my thanks to Professor I. Ojima for providing me with some information I should have known.
\normalsize

\section{Measurement apparatus}
In the standard setting originated from von Neumann's book \cite{vN}, a {\em measuring process} for the quantum system on a separable Hilbert space $\Hil$ is described by a quadruple $(\cL,\phi,M,U)$, where $\cL$ is a separable Hilbert space, $\phi$ is a state\footnote{A state $\phi$ of $\K(\cL)$ corresponds to a positive trace class operator $\rho$ on $\cL$ of trace one; $\phi(x)=\Tr(x\rho),\ x\in \K(\cL)$.} of the compact operators $\K(\cL)$, $M$ is a self-adjoint operator on $\cL$, and $U$ is a unitary on $\Hil\otimes\cL$.\footnote{Taken from \cite{Oz} with some notational modifications.} Let $E_M$ denote the spectral measure of $M$ and let $\K(\Hil)^*_+$ denote the cone\footnote{This is the same as the cone of positive trace class operators on $\Hil$.} of positive functionals on $\K(\Hil)$ and $\cS(\K)=\{\varphi\in \K(\Hil)^*_+\ |\ \|\varphi\|=1\}$, the convex state space of $\K=\K(\Hil)$. This process produces an $\E(\Delta,\varphi)\in \K^*_+$ for each Borel subset $\Delta$ of $\R$ and $\varphi\in \cS(\K)$ by
$$
\E(\Delta,\varphi)(x)=\varphi\otimes\phi(U^*(x\otimes E_M(\Delta))U),\ \ x\in \K.
$$
Note that $\Delta\mapsto \E(\Delta,\varphi)$ is a $\K^*_+$-valued measure on $\R$ with $\E(\R,\varphi)\in \cS(\K)$ and that $\varphi\mapsto \E(\Delta,\varphi)$ is a continuous affine map from $\cS(\K)$ into $\K^*_+$ (in norm). Note also that if $(\Delta_n)$ is an increasing sequence of Borel subsets of $\R$ then $\E(\Delta_n,\varphi)$ converges to $\E(\bigcup_n\Delta_n,\varphi)$. All what we get from the quadruple $(\cL,\phi,M,U)$ with $U$ a unitary on $\Hil\otimes\cL$ is the collection $\E(\Delta,\varphi)$, which is called a {\em Davies-Lewis instrument} or DL-instrument for short \cite{DL70}. The self-adjoint operator $M$ is called a {\em meter observable} in \cite{Oz} and a {\em probe observable} in \cite{Oz03}; When the observed system $\K(\Hil)$ is under the state $\varphi$ and is applied this measuring process, we are supposed to be able to perform a {\em precise local measurement}\footnote{It is a bit bizarre to assume "`measurement"' in some sense is possible at all to explain measurement.}  of $M$ to get a real number $x$ whose distribution is given by $\Delta\mapsto \E(\Delta,\varphi)(1)$ so that the ensemble of all states of $\K(\Hil)$ after observing $x\in\Delta$ is given by $\E(\Delta,\varphi)/\E(\Delta,\varphi)(1)$ for each $\Delta$ (cf. Section 3 of \cite{Oz84}). Given a self-adjoint operator $Q$ on $\Hil$, observing $Q$ is supposed to be choosing a suitable $(\cL,\phi,M,U)$ and applying the above process.

This seems to be all well-established except for how to drive the quantum effect on $M$ to the macroscopic level for observation (cf. \cite{Oz97}). Recently Harada and Ojima describe such an amplification process as well as the preceding interaction between the observed and the probe systems in terms of certain abelian groups by noting a specific property of {\em regular representation} (section 3 of \cite{Oj09}). Here we propose another mathematical model for measurements of a quantum system in a C*-algebra setting, which incorporates a mechanism of {\em magnifying quantum effects to the classical level}\footnote{This expression is taken from page 250 of Penrose's book \cite{Pen}.} into the measuring apparatus. In this scheme the state of the quantum system transforms to new ones according to a certain probability law just like the phase does to a new one in phase transition we encounter in equilibrium quantum statistical mechanics.

The (microscopic) quantum system is described by the C$^*$-algebra $\K=\K(\Hil)$ of compact operators on a separable Hilbert space $\Hil$ just as above and the (macroscopic) measuring apparatus is by a unital separable non-type I nuclear simple C$^*$-algebra $A$ with a certain unital endomorphism and a pure state.\footnote{That the C$^*$-algebra $A$ is non-type I and nuclear is assumed to assure that $A$ has a desired endomorphism \cite{K03}. We may assume that $A$ is the UHF algebra of type $2^\infty$, or the infinite tensor product of $2\times 2$ matrices, which may be considered as the observable algebra for electrons with an infinite degrees of freedom.} We will then specify a unitary from the combined system $\K\otimes A$ to dictate an interaction. After applying the adjoint action of the unitary and the endomorphism we reach a situation similar to the above; instead of $M$ (or the von Neumann algebra generated by $M$) we will obtain an abelian von Neumann algebra, as the center of the observable algebra, as an outcome of this process.

We have introduced $FS(A)$ and $FR(A)$. Let $S(A)$ denote the set of all states of $A$. Then $S(A)$ is the closed convex subset of $A^*$ and let $PS(A)$ be the set of extreme points of $S(A)$ each of which is called a pure state. Then the GNS representation $\pi_\omega$ associated with $\omega\in PS(A)$ are irreducible; hence $PS(A)\subset FS(A)$. We will later concentrate on $PS(A)$ instead of $FS(A)$ because our knowledge on endomorphisms is limited.

If $\gamma$ is a unital endomorphism of $A$ with $\gamma(A)\not=A$ and $\phi$ is a factorial state then $\phi\gamma$ is a state but may not be factorial, i.e., if $\M$ denotes the weak closure of $\pi_{\phi\gamma}(A)$, the center $\M\cap \M'$ of $\M$ may not be $\C1$. (If $\phi\gamma$ is factorial $\pi_\phi\gamma$ may not be factorial, i.e., the weak closure of $\pi_\phi\gamma(A)$ may have non-trivial center. This is because $\pi_{\phi\gamma}$ is the restriction of $\pi_\phi\gamma$ to the subspace defined as the closure of $\pi_\phi\gamma(A)\Omega_\phi$.) In this case $\phi\gamma$ is centrally decomposed in the sense that there is a unique probability measure $\nu$ on the Borel subset $FS(A)$ of factorial states in $A^*$ (under the assumption that $A$ is separable) with
$$
\phi\gamma=\int_\F \psi {\rm d}\nu(\psi),
$$
where $\M\cap \M'$ on the left could be identified with $L^\infty(\F,\nu)$ on the right behind this equality (see 3.1.8 and 3.4.5 of \cite{Sak}). If $\phi\in FS(A)$ transforms to $\phi\gamma$ causally but in a irreversible way then it would immediately jump to $\psi\in FS(A)$ acausally according to the probability $\nu$ on $FS(A)$.\footnote{Phases or sectors are also discussed in \cite{Oj09}. See \cite{BR} for backgrounds.}

We also assume that $\gamma$ is asymptotically inner.\footnote{This is a misnomer but is widely used among operator algebraists. It is more like being asymptotically NOT inner and means that $\gamma$ is asymptotically approximated by inner automorphisms.} Namely we assume that there is a continuous unitary path $u_t,\,t\in [0,1)$ in $A$ such that $\gamma(x)=\lim_{t\rightarrow1}u_txu_t^*,\ x\in A$ and $u_0=1$. Then it follows that there is a bounded sequence $(h_n)$ of self-adjoint elements of $A$ such that $\lim_n[h_n,x]=0$ and $\gamma(x)=\lim_n \Ad(e^{ih_1}e^{ih_2}\cdots e^{ih_n})(x)$ for $x\in A$. We regard $\gamma$ as a time development as being a limit of Hamiltonean induced time developments which are cascading quantum effects to the visible classical level within a small time interval. Thus $\gamma$ describes an irreversible process.\footnote{An ideal measuring apparatus, interacting with the quantum system, should never be interfered by external forces but has to yield classical information to the outside. This is different from a closed system whose time evolution is described by a group of automorphisms and may be called a decaying closed system (presumably not sustainable indefinitely). It is not an open system either, which is ideally described by a semigroup of CP contractions incorporating external forces.}

If $\varphi$ is a state of $\K(\Hil)$ and the measuring apparatus $A$ is in a pure state $\phi$, then we suppose that $\varphi\otimes\phi$ turns to $(\varphi\otimes\phi)\Ad\,U^*$ and then to $(\varphi\otimes\phi)\Ad\,U^*(\id\otimes \gamma)$, which may not be factorial and then will be centrally decomposed as explained above.

We formally give the definition of DL instrument or rather CP instrument following \cite{Oz84} and then the definition of our measuring processes (cf. \cite{DL70,Oz84,Oz}).

\begin{definition}\label{I}
Let $\M$ be an abelian von Neumann algebra with separable predual\footnote{A Banach space $X$ is a predual of $\M$ if $X^*\cong \M$; $\M$ has a unique predual denoted by $\M_*$ (1.13.3 of \cite{Sak}). We know that $\M$ is isomorphic to the $L^\infty$-space on some probability space.}  and $\M_+$ the cone of positive elements of $\M$. Let $\Hil$ be an infinite-dimensional separable Hilbert space and $\K=\K(\Hil)$ be the C$^*$-algebra of compact operators on $\Hil$.  We call a map $\E$ from $\M\times \K^*$ into $\K^*$ a {\em CP instrument} based on $\M$ if it satisfies
\begin{enumerate}
\item For each $\varphi\in \K^*_+$ the map $\M\ni Q\mapsto \E(Q,\varphi)\in \K^*$ is a positive continuous linear map on $\M$,
\item For each $Q\in\M_+$ the map $\K^*\ni\varphi\mapsto \E(Q,\varphi)\in \K^*$ is a completely positive (CP for short) linear map,
\item $\E(1,\varphi)(1)=\varphi(1)$ for all $\varphi\in \K^*$,
\end{enumerate}
where $\M$ is equipped with the weak$^*$-topology.
\end{definition}

If we denote by $\E(Q)$ the linear map $\K^*\ni\varphi\mapsto \E(Q,\varphi)\in \K^*$ with $Q\in \M_+$, the dual map $\E(Q)^*: \K^{**}=\B(\Hil)\ra \B(\Hil)$ is completely positive or CP, i.e., the natural extension of $\E(Q)^*$ to a map from $M_k\otimes \B(\Hil)$ into $M_k\otimes \B(\Hil)$ is positive for any $k\in\N$, which follows from the complete positivity of $\E(Q)$. We denote $\E(Q)^*b$ by $\E^*(Q,b)$ for $b\in \B(\Hil)$; then for each $b\in \B(\Hil)_+$ the map $Q\mapsto \E^*(Q,b)$ is a positive continuous linear map\footnote{Since $\M$ is commutative this map is automatically CP.} when $\M$ and $\B(\Hil)$ are endowed with the weak$^*$-topology. For $Q\in \M_+$ the map $b\mapsto \E^*(Q,b)$ is a CP continuous linear map when $\B(\Hil)$ is endowed with the weak$^*$-topology. The third condition of the above definition is equivalent to $\E^*(1,1)=1$.

\begin{definition}\label{D}
Let $A$ be a unital separable non-type I nuclear simple C$^*$-algebra. Let $\phi$ be a pure state of $A$ and $\gamma$ an asymptotically inner endomorphism of $A$ such that $\pi_\phi\gamma(A)'$ is a non-trivial abelian von Neumann algebra. Let $\K=\K(\Hil)$ be as in the above definition and let $U$ be a unitary in the multiplier algebra $M(\K\otimes A)$\footnote{Identifying $\K\otimes A$ with $\K\otimes \pi_\phi(A)$ on $\Hil\otimes\Hil_\phi$, the multiplier algebra $M(\K\otimes A)$ is the set of $Q\in \B(\Hil\otimes\Hil_\phi)$ satisfying $Q(\K\otimes A),(\K\otimes A)Q\subset \K\otimes A$. For any unitary $U\in M(\K\otimes A)$ there is a unitary path $U_t,\ t\in [0,1]$ in $M(\K\otimes A)$ such that $U_0=1$, $U_1=U$, and $t\mapsto xU_t, U_tx$ are continuous in $\K\otimes A$ (12.2.2 of \cite{Bl}). Thus we may regard $\Ad\,U^*$ as representing a time development of $\K\otimes A$.} of $\K\otimes A$. We call $(A,\phi,\gamma,U)$ a {\em measuring process} for $\K$.\addednote{We may take a separable C$^*$-algebra $B$ for the observed system instead of $\K$. Then a measuring process for $B$ is defined as $(A,\phi,\gamma,U)$ where $U$ is now a unitary $U$ of $M(B\otimes A)$ connected with $1$.}
\end{definition}

\begin{prop}\label{CP}
Let $(A,\phi,\gamma,U)$ be a measuring process and let $\M=\pi_\phi\gamma(A)'$. For each $Q\in\M$ and $\varphi\in \K^*$  define an $\E(Q,\varphi)\in\K^*$ by
$$
\E(Q,\varphi)(x)=\overline{\varphi\otimes\phi}(\Ad\,\bar{U}^*)(x\otimes Q),\ \ x\in \K,
$$
where $\overline{\varphi\otimes\phi}$ is a unique extension of the positive functional $\varphi\otimes\phi\pi_\phi^{-1}$ of $\K\otimes\pi_\phi(A)$ to a weak$^*$-continuous one of $(\K\otimes\pi_\phi(A))''=\B(\Hil\otimes\Hil_\phi)$ and $\bar{U}=(\id\otimes\pi_\phi)(U)$. Then $\E$ is a CP instrument based on $\M$. We call $\E$ the {\em CP instrument} obtained from $(A,\phi,\gamma,U)$.\addednote{If the observed system is a general separable C$^*$-algebra $B$, we specify an irreducible representation $\pi$ of $B$ and denote by $V_\pi$ as the linear space consisting of $\varphi\in B^*$ such that $\varphi=f\pi$ for some $f\in \pi(B)''_*$. Then as $V_\pi^*=\pi(B)''=\B(\Hil_\pi)$ each $\pi$ and a measuring process $(A,\phi,\gamma,U)$ for $B$ define a CP instrument $\E(\varphi,Q)\in V_\pi$ for $\varphi\in V_\pi, Q\in \M$ just as above. If $B=\K$ then there is essentially only one $\pi$.}
\end{prop}

If $E_\phi$ denotes the conditional expectation of $\B(\Hil\otimes \Hil_\phi)$ onto $\B(\Hil)$ defined by
$$
\varphi(E_\phi(T))=\overline{\varphi\otimes \phi}\Ad \bar{U}^*(T),\ \ \varphi\in \K(\Hil)^*=\B(\Hil)_*
$$
then it follows that $\E(Q,\varphi)(1)=\varphi(E_\phi(1\otimes Q))$. If $E_\phi|\M$ is a homomorphism, one can say that $(A,\phi,\gamma,U)$ exactly observes the abelian von Neumann algebra $E_\phi(1\otimes\M)$ (or a self-adjoint operator which generates it). In general it does only approximately an observable residing in $\Hil$.

Note that we only use $\M=\pi_\phi\gamma(A)'$ for construction of the CP instrument $\E$, not $\gamma$ directly. In this sense the present scheme is not much different from the original one by von Neumann on the technical level. But we hope that the present model makes a contribution to a clarification on the conceptual level.

When $\E_1$ and $\E_2$ are CP instruments based on $\M$ and $(\xi_n)$ is a dense sequence in the unit sphere of $\Hil$ we define  $d(\E_1,\E_2)\geq 0$ by
 $$
 d(\E_1,\E_2)=\sum_n2^{-n} \|\E_1(\ \cdot\ ,\psi_n)-\E_2(\ \cdot\ ,\psi_n)\|,
 $$
where $\psi_n$ is the vector state of $\K$ defined by $\xi_n$. It follows that $d$ is a distance on the set of CP instruments based on $\M$. We can show the following:

\begin{prop}
Let $\M$ be an abelian von Neumann algebra with separable predual. Then in the set of all  CP instruments based on $\M$ is dense the set of CP instruments obtained from the measuring processes $(A,\phi,\gamma,U)$ with $\M=\pi_\phi\gamma(A)'$ in the sense of Proposition \ref{CP}.
\end{prop}

We will sketch how to prove this. First of all we have to show that there is an asymptotically inner endomorphism $\gamma$ and an irreducible representation $\pi$ of some unital separable non-type I nuclear simple C$^*$-algebra $A$ such that $\M\cong \pi\gamma(A)'$ (or $\pi\gamma(A)''\cong \M\otimes \B(\Hil_1)$ where $\Hil_1$ is an infinite-dimensional separable Hilbert space). This is indeed possible for any unital separable non-type I nuclear C$^*$-algebra, whose proof requires Glimm's result \cite{Gl} (which shows UHF algebras are typical examples of non-type I C$^*$-algebras), the existence result on endomorphisms \cite{K03} (for non-type I nuclear C$^*$-algebras), and the following well-known statement on UHF algebras: For any such $\M$ as above there is a representation $\pi$ such that $\pi(A)'\cong \M$, which will be shown in the same way as the examples of endomorphisms are given in Section 3. Thus we prepare $(A,\gamma)$ and some irreducible representation $\pi$ with $\pi\gamma(A)'\cong \M$.

Secondly by Ozawa's results (5.1-3 of \cite{Oz84}) all the CP instruments are realized by the measuring processes in his sense (stated in the beginning). For the proof we use the fact that $\M\times \B(\Hil)\ni (Q,b)\mapsto \E^*(Q,b)\in \B(\Hil)$ is a completely positive, weak$^*$-continuous bilinear map and express this map as the restriction of a {\em faithful} weak$^*$-continuous representation of $\M\otimes \B(\Hil)$ (by extending if necessary the representation obtained by Stinespring's
procedure as in the proof of 4.2 of \cite{Oz84}).  Namely for a CP instrument $\E$ based on $\M$ one finds a separable Hilbert space $\cL$, a pure state $\phi$ of $\K(\cL)$, a normal unital representation $\rho$ of $\M$ on $\cL$, and a unitary $U$ on $\Hil\otimes\cL$ such that
$$
\E(Q,\varphi)(x)=\overline{\varphi\otimes\phi}(\Ad\,U^*(x\otimes\rho(Q)),\ \ Q\in \M,  \ x\in\K.
$$
We may assume that $\rho(\M)'\cong \rho(\M)\otimes \B(\Hil_1)$ with $\dim(\Hil_1)=\infty$ by tensoring $\cL$ by another infinite-dimensional separable Hilbert space if necessary and making obvious arrangements. Then we outfit an irreducible representation $\pi$ of $A$ on $\cL$ such that $\pi\gamma(A)'=\rho(\M)$.

Since this is done independently of $U$, we cannot expect that $U\in M(\K\otimes \pi(A))$. But, noting that $(\K\otimes \pi(A))''=\B(\Hil\otimes\cL)$, Kadison's transitivity (\cite{Kad} or 1.21.16 of \cite{Sak}) tells us that one can find a unitary $u\in \K\otimes \pi(A)+\C1$ which equals $U$ on any given finite-dimensional subspace.\footnote{Which shows a slightly stronger statement: For any CP instrument $\E$ and any finite number of pure states $\varphi_1,\ldots,\varphi_n$ on $\K$ there is a measuring process whose CP instrument is equal to $\E$ on $\varphi=\varphi_1,\ldots,\varphi_n$ (and any $Q\in\M$).}  Thus we can replace $U$ by a unitary in $M(\K\otimes A)$ so that the resulting CP instrument is arbitrarily close to $\E$.

In the next section we will give an example of measuring process and explain the above definition of CP instruments in more details. In section 3 we will show how to construct endomorphisms and irreducible representations in the case of UHF algebras of type $k^\infty$. I wonder if this exposition gives some justification for $\gamma$ being a magnifying glass of quantum effects.

\section{The case $\pi\gamma(A)'\cong \ell^\infty(\N)$}

Let $A$ be a unital separable non-type I nuclear simple C$^*$-algebra and let $\phi$ be a pure state of $A$.  Let $\gamma$ be an asymptotically inner endomorphism of $A$ such that $\pi_\phi\gamma(A)'$ is an arbitrary abelian von Neumann algebra. The existence of such $\gamma$ follows from Theorem 3.3 of \cite{K03}.\footnote{For example let $\nu_1,\nu_2,\ldots,$ be a sequence of irreducible representations of $A$ such that all $\nu_n$ are mutually disjoint. If $\rho$ is the direct sum $\nu_1\oplus\nu_2\oplus\cdots$ then the weak closure of $\rho(A)$ is equal to $\B(\Hil_1)\oplus \B(\Hil_2)\oplus\cdots$, where $\Hil_n$ is the representation space of $\nu_n$. Then by Theorem 3.3 of \cite{K03} it follows that there is an asymptotically inner endomorphism $\gamma$ of $A$ such that $\pi\gamma$ is unitarily equivalent to $\rho$, which implies that $\pi\gamma(A)'$ is isomorphic to $\C\oplus\C\oplus\cdots$. The condition $u_0=1$ for the choice of $u_t$ is not explicitly mentioned but follows from the proof. We could impose a finite number of conditions on $\gamma$ of similar nature $\pi_i\gamma\cong \rho_i$ with mutually disjoint irreducible $\pi_i$ and arbitrary $\rho_i$.} Let $U$ be a unitary in $M(\K\otimes A)$. We will describe how the system $(A,\phi,\gamma,U)$ works as a measuring apparatus for the observed quantum system $\K$.

Let $\varphi$ be a state of $\K$. We denote by $\id$ the identity representation of $\K=\K(\Hil)$ on $\Hil$, where $\varphi$ extends to a normal state of $\B(\Hil)=\K(\Hil)''$. Then through the interaction with $(A,\phi)$ the state $\varphi\otimes\phi$ of the combined system $\K\otimes A$ changes to $(\varphi\otimes\phi)\Ad\,U^*$, and then to $T(\varphi)=(\varphi\otimes \phi)\Ad\,U^*(\id\otimes \gamma)$. Let $\pi_0=(\id\otimes\pi_\phi)\Ad\,U^*(\id\otimes \gamma)$, which is a representation of $\K\otimes A$ on the Hilbert space $\Hil\otimes\Hil_\phi$. Then the commutant $\pi_0(\K\otimes A)'$ is equal to $\Ad\,\bar{U}^*(\C1\otimes \pi_\phi\gamma(A)')$, where $\bar{U}=\id\otimes\pi_\phi(U)$. Note that $\pi_0(\K\otimes A)'=\pi_0(\K\otimes A)'\cap \pi_0(\K\otimes A)''$, the center of $\pi_0(\K\otimes A)''$.

Suppose that $\pi_\phi\gamma(A)'\cong \ell^\infty(\N)$, i.e., it is generated by minimal projections $E_1,E_2,\ldots$ on $\Hil_\phi$. Since $x\mapsto\pi_\phi\gamma(x)E_i$ is an irreducible representation of $A$ on $E_i\Hil_\phi$, $E_i$ is of infinite rank. Let $F_i=\Ad\,\bar{U}^*(1\otimes E_i)$, which is a minimal projection of the center of $\pi_0(\K\otimes A)''$. If $\overline{\varphi\otimes \phi}$ denotes the natural extension to a normal state of $\B(\Hil\otimes \Hil_\phi)$ then
$$
T(\varphi)=\sum_{i=1}^\infty \overline{\varphi\otimes \phi}(F_i\pi_0(\ \cdot\ )).
$$
Since $F_i$ is a minimal projection in $\pi_0(\K\otimes A)'$, the state $\omega_i=\overline{\varphi\otimes\phi}(F_i\pi_0(\ \cdot\ ))/\overline{\varphi\otimes\phi}(F_i)$ is a pure state of $\K\otimes A$ for $\overline{\varphi\otimes\phi}(F_i)>0$. Since $F_i$'s are mutually orthogonal central projections,  $\omega_i$'s are mutually disjoint.\footnote{$\omega_1$ and $\omega_2$, states of $B=\K\otimes A$, are disjoint if and only if there is a central sequence $(x_n)$ in $B$ such that $\omega_1(x_n)\ra1$ and $\omega_2(x_n)\ra0$, which implies that $(\pi_{\omega_1}\oplus \pi_{\omega_2})(x_n)\ra 1\oplus 0$ in the weak operator topology. $(x_n)$ is a central sequence if it is bounded and $[x_n,y]\ra0$ for any $y\in B$. The C$^*$-algebra consisting of central sequences (divided by some trivial ones) is considered to be the classical observables associated with $B$. We expect they reduce to numbers in a phase.} Hence $T(\varphi)$ is the sum of phases with weights and Nature will pick up one according to the probability specified by $(\overline{\varphi\otimes \phi}(F_i))$.

Since $\varphi\mapsto \overline{\varphi\otimes\phi}(F_i)$ extends to a continuous positive linear map from $\K^*$ into $\C$ there is a positive operator $P_i$ in $\B(\Hil)=\K(\Hil)^{**}$ such that $\varphi(P_i)=\overline{\varphi\otimes\phi}(F_i)$. Since $\sum_iF_i=1$ it follows that $\sum_iP_i=1$. Note that restriction of $\overline{\varphi\otimes\phi}(F_i\pi_0(\ \cdot\ ))$ to $\K$ is $\E(E_i,\varphi)=\overline{\varphi\otimes\phi}\Ad\,\bar{U}^*(\ \cdot\ \otimes E_i)$ and $\varphi(P_i)=\E(E_i,\varphi)(1)$ using the notation given in Definition \ref{D}.

Hence it follows that
$$
T(\varphi)|\K=\sum_i\varphi(P_i)\frac{\E(E_i,\varphi)}{\varphi(P_i)},
$$
where the sum is over $i$ with $\varphi(P_i)>0$ and $\varphi_i=\E(E_i,\varphi)/\varphi(P_i)$ is a state of $\K$, not necessarily a pure state. Here is our conclusion: After applying this measuring process to $\K$ Nature will transform $\varphi$ to $\varphi_i$ with probability $\varphi(P_i)$ for each $i=1,2,\ldots$.

Note that if $U=1$ then $P_i=\overline{\varphi\otimes\phi}(1\otimes E_i)1$ is independent of $\varphi$. Suppose that $\phi$ is given as a vector state by a unit vector $\psi_1\in E_1\Hil_\phi$. If $U=1$ then $T(\varphi)=\varphi\otimes \phi\gamma$ is pure and $P_1=1$ (and other $P_i=0$); no information is gained. We choose a unit vector $\psi_i\in E_i\Hil_\phi$ for each $i>1$ and choose a unitary $u_i\in A$ (or $A+\C1$ if $A$ is non-unital) for $i\geq 1$ such that $\pi_\phi(u_i)\psi_1=\psi_i$. The existence of such $u_i$ follows from Kadison's transitivity \cite{Kad} since $\pi_\phi$ is irreducible. We set $U=\sum_ie_{ii}\otimes u_i$; the summation converges to a unitary as a multiplier of $\K\otimes A$, where $(e_{ij})$ are matrix units generating $\K$. Since $\bar{U}\xi_i\otimes \psi_1=\xi_i\otimes \psi_i$ where $(\xi_i)$ is an orthnormal basis of $\Hil$ with $e_{ii}\xi_i=\xi_i$, it follows that $\overline{\varphi\otimes \phi}(F_i)=\overline{\varphi\otimes\phi}(\bar{U}^*(1\otimes E_i)\bar{U})=\varphi(e_{ii})$ and $\varphi_i(x)=\Tr(e_{ii}x)$ for $x\in \K$ (when $\varphi(e_{ii})>0$). Hence for this choice of $\phi$ and $U$ we obtain
$$
T(\varphi)|\K=\sum_i\varphi(e_{ii})\Tr(e_{ii}\ \cdot\ ),
$$
which is what we expect by measuring e.g., the unbounded observable $\sum_n ne_{nn}\in M(\K)$.

We should note that the von Neumann algebra $\M$ generated by all $E_i$ plays the same role as the von Neumann algebra generated by $M$ for the measuring process $(\cL,\phi,M,U)$ with $\cL=\Hil_\phi$ we discussed in the beginning. Previously $M$ is just an arbitrary self-adjoint operator on $\Hil_\phi$ and so the von Neumann algebra generated by $M$ can contain a non-zero compact operator. But the present $\M$ must satisfy $\M\cap \K(\Hil_\phi)=\{0\}$.\footnote{But this is not important as it is attained by tensoring an infinite-dimensional Hilbert space.}

\section{Endomorphisms}
As we have noted, Theorem 3.3 of \cite{K03} serves to produce the desired endomorphisms for any separable non-type I nuclear simple C$^*$-algebra. Here we show a concrete way to construct an asymptotically inner endomorphism $\gamma$ and an irreducible representation $\pi$ for the UHF algebra $A$ of type $k^\infty$ with $k>1$ such that $\pi\gamma(A)'$ is isomorphic to $\C^k$ but $A\cap \gamma(A)'=\C1$.\footnote{Which I do not have a specific reason to require but consider as a condition for $\gamma$ to be close to an automorphism.}

We denote by $M_k$ the C$^*$-algebra of $k\times k$ matrices and denote by $v$ the diagonal matrix $1\oplus \omega\oplus \omega^2\oplus\cdots\oplus \omega^{k-1}$ with $\omega=e^{i2\pi/k}$. We define an automorphism $\sigma$ on $A=M_k\otimes M_k\otimes \cdots$ by
$$
\sigma=\bigotimes_{n=1}^\infty\Ad\,v.
$$
Since $v^k=1$ it follows that $\sigma^k=\id$. The fixed point algebra $A^\sigma=\{x\in A\ |\ \sigma(x)=x\}$ is isomorphic to $A$ (see \cite{St70} and \cite{K77}). This is easy to see if you know of AF algebras and associated Bratteli diagrams \cite{Br72}. We regard  $M_k^{\otimes n}$ as the C$^*$-subalgebra of $A$ generated by the first $n$ copies of $M_k$.
Since $v^{\otimes n}=\sum_{j=0}^{k-1}\omega^j E_j\in M_k^{\otimes n}$, where $E_j$'s are mutually orthogonal projections of $M_k^{\otimes n}$ of rank $k^{n-1}$, it follows that $(M_k^{\otimes n})^\sigma=\bigoplus_{j=0}^{k-1}E_jM_k^{\otimes n}E_j$ with $E_jM_k^{\otimes n}E_j\cong M_k^{\otimes (n-1)}$. Thus we can embed $M_k^{\otimes (n-1)}$ into $(M_k^{\otimes n})^\sigma$. We construct such embeddings consistently from $M_k\subset M_k^{\otimes 2}\subset M_k^{\otimes 3}\subset \cdots$ into $(M_k^{\otimes 2})^\sigma\subset (M_k^{\otimes 3})^\sigma\subset (M_k^{\otimes 4})^\sigma\subset\cdots$ preserving each level, where the closure of the union of the former (resp. latter) sequence is $A$ (resp. $A^\sigma$); thus we obtain the isomorphism $\gamma$ of $A$ onto $A^\sigma$, which is the endomorphism we aimed at and is asymptotically inner as all the unital endomorphisms of $A$ are. In our case this is also easy to see. Since $\gamma(M_k)\subset M_k^{\otimes 2}$, there is a (continuous) unitary path $u_t^{(1)},\ t\in [0,1]$ in $M_k^{\otimes 2}$ such that $u^{(1)}_0=1$ and $\Ad\,u^{(1)}_1(M_k)=\gamma(M_k)$. Since $\Ad\,u^{(1)}_1(M_k^{\otimes 2})=\gamma(M_k)\Ad\,u_1^{(1)}(1\otimes M_k)\subset M_k^{\otimes 3}$, it follows that both $\Ad\,u^{(1)}_1(1\otimes M_k)$ and $\gamma(1\otimes M_k)$ are unital subalgebras of $M_k^{\otimes 3}\cap \gamma(M_k)'$. Hence there is a unitary path $u_t^{(2)},\ t\in [0,1]$ in $M_k^{\otimes 3}\cap \gamma(M_k)'\cong M_k\otimes M_k$ such that $u_0^{(2)}=1$ and $\Ad(u_1^{(2)}u^{(1)}_1)(1\otimes M_k)=\gamma(1\otimes M_k)$. In this way we construct a unitary path $u_t^{(n)},\ t\in [0,1]$ in $M_k^{\otimes (n+1)}\cap \gamma(M_k^{(n-1)})'$ such that $u_0^{(n)}=1$ and $\Ad(u^{(n)}_1u^{(n-1)}_1\cdots u_1^{(1)})(1^{\otimes (n-1)}\otimes M_k)=\gamma(1^{\otimes (n-1)}\otimes M_k)$. Combining all these unitary paths $u^{(n)}_t$ into one continuous unitary path $u_t,\ t\in [0,\infty)$ in $A$ it follows that $\gamma(x)=\lim_{t\ra\infty}\Ad\,u_t(x),\ x\in A$. Thus $\gamma$ is an asymptotically inner endomorphism such that $\gamma(A)=A^\sigma$.

Let $\phi_0$ be the pure state of $M_k$ defined by $\phi_0(x)=x_{11}$ for $x=(x_{ij})\in M_k$. Since $\phi_0(v)=1$ we have that $\phi_0\Ad\,v=\phi_0$. Let $\phi=\phi_0\otimes \phi_0\otimes\cdots$, which is a $\sigma$-invariant pure state of $A$. Define a unitary $U$ on the GNS representation space $\Hil_\phi$ by
$$
U\pi_\phi(x)\Omega_\phi=\pi_\phi\sigma(x)\Omega_\phi,\  x\in A.
$$
Then $\Ad\,U\pi_\phi(x)=\pi_\phi\sigma(x),\ x\in A$ and $U^k=1$. We can conclude that $\pi_\phi\gamma(A)'= U''\cong \C^k$ and that $\phi|\gamma(A)$ is a pure state.

By using the above fact we can construct more general examples. Since the tensor product of an infinitely many copies of $A$ is isomorphic to $A$, one obtains a unital endomorphism of $A$ by $A\cong A\otimes A\otimes\cdots\ra \gamma(A)\otimes\gamma(A)\otimes\cdots \subset A\otimes A\otimes\cdots\cong A$, where the middle map is defined by $\gamma\otimes\gamma\cdots$. We denote this unital endomorphism by $\gamma^\infty$. Let $\phi^\infty$ be the pure state of $A\cong A\otimes A\otimes\cdots$ defined by $\phi\otimes\phi\otimes\cdots$. Since $\phi^\infty$ is invariant under the action $\sigma^\infty=\sigma\otimes\sigma\otimes\cdots$ of the compact group $G=\Z_k\times\Z_k\times\cdots$ on $A\cong A\otimes A\otimes\cdots$ and $\phi^\infty|\gamma^\infty(A)$ is pure, it follows that $\pi_{\phi^\infty}\gamma^\infty(A)'$ is isomorphic to $\ell^\infty(\hat{G})\cong\ell^\infty(\N)$.

Define a unit vector $\xi\in \Hil_\phi$ by $\xi=n^{-1/2}\sum_{j=1}^k\pi_\phi(e_{j1}^{(1)})\Omega_\phi$ and a pure state $\psi$ of $A$ by $\psi(x)=\lan \xi, \pi_\phi(x)\xi\ran$, where $(e^{(1)}_{ij})$ is the matrix units of $M_k\subset A$. Then $\psi$ is not $\sigma$-invariant but $\sigma$-covariant (i.e., $\pi_\psi=\pi_\phi$ is $\sigma$-covariant). Let $\psi^\infty$ is the state of $A\cong A\otimes A\otimes \cdots$ defined by $\psi\otimes\psi\otimes\cdots$. Then $\psi$ is covariant under the action obtained by restricting $\sigma^\infty$ to the discrete subgroup $\hat{G}=\bigcup_{n=1}^\infty \Z_k\times \Z_k\times\cdots \times\Z_k (n\ {\rm factors})$ of $G$. Since there are no $\sigma^\infty|\hat{G}$-invariant states associated with $\pi_{\psi^\infty}$ we can conclude that $\pi_{\psi^\infty}\gamma(A)'\cong L^\infty(G)$, which is completely non-atomic. (If $\pi_{\psi^\infty}\gamma(A)'$ has a minimal projection $E$ then a unit vector in $E\Hil_{\psi^\infty}$ defines a $\sigma^\infty|\hat{G}$-invariant state of $A$, which must be $\sigma^\infty$-invariant, leading us to a contradiction.)

Let $\phi_1$ be the pure state of $M_k$ defined by $\phi_1(x)=k^{-1}\sum_{i,j}x_{ij}$ and let $\chi=\phi_1\otimes\phi_1\otimes\cdots$ as a state of $A$. We define $\gamma$ in the most natural way, i.e., regarding $M_k^{\otimes (n-1)}=M_k(M_k^{\otimes (n-2)})$ as a giant matrix algebra $M_{k^{(n-1)}}$ in the natural way we embed $M_k^{\otimes (n-1)}$ into $(M_k^{\otimes n})^\sigma$ componentwise. Then one can easily see that $\chi\gamma=\chi$. Since $\chi,\chi\gamma,\cdots,\chi\gamma^{k-1}$ are mutually disjoint, it follows that $\pi_\chi\gamma(A)'=\C1$ as $\gamma(A)=A^\sigma$. In particular $A\cap\gamma(A)'=\C1$. In the representation $\pi_\chi$ there is a unitary $U$ such that $\Ad\,U\pi_\chi(x)=\pi_\chi\gamma(x),\ x\in A$, i.e., $\gamma$ is at least implemented by a unitary in some representation like an automorphism.

\end{document}